\documentclass[journal]{IEEEtran}

\usepackage{algorithm,algpseudocode}
\usepackage{amstext,amsmath,amssymb,exscale,mathtools,bm,gensymb,cite}
\usepackage{psfrag}
\usepackage[none]{hyphenat}
\usepackage{tabularx}
\usepackage[shortlabels]{enumitem}
\usepackage{xcolor}
\usepackage[font=small]{caption}
\usepackage{subcaption,siunitx}
\usepackage{scalerel}
\usepackage{makecell}
\usepackage{graphicx}
\usepackage[nolist]{acronym}   
\usepackage{hyperref}

\begin{document}

\renewcommand{\baselinestretch}{0.90}

\begin{acronym}[MPC] 
	
	\acro{6G}{sixth generation} 
	
	\acro{AR}{auto-regressive}
	
	\acro{AoA}{angle of arrival}
	
	\acro{AoD}{angle of departure}
	
	\acro{AWGN}{additive white Gaussian noise}
	
	\acro{B5G}{beyond fifth generation}
	
	\acro{BS}{base station}
	
	\acro{BPSK}{binary phase-shift keying}
	
	\acro{IRS}{intelligent reflecting surface}
	
	\acro{MIMO}{multiple input multiple output}
	
	\acro{CSI}{channel state information}
	
	\acro{DFT}{discrete Fourier transform}
	
	\acro{MMSE}{minimum mean squared error}
	
	\acro{NMSE}{normalized mean squared error}
	
	\acro{SNR}{signal-to-noise ratio}
	
	\acro{SVD}{singular value decomposition}
	
	\acro{SE}{spectral efficiency}
	
	\acro{UE}{user equipment}
	
	\acro{ULA}{uniform linear array}
	
	\acro{URA}{uniform rectangular array}
	
	
	\acro{ALS}{alternating least squares}
	
	\acro{BALS}{bilinear alternating least squares}
	
	\acro{KF}{Kronecker factorization}
	
	\acro{KRF}{Khatri-Rao factorization}
	
	\acro{LS}{least squares}    
	
	\acro{PARKRON}{PARAFAC-Khatri-Rao-Kronecker factorization}
	
	\acro{TBT}{Tucker-based tracking}
	
\end{acronym}


\title{Tensor-Based Channel Estimation and Data-Aided Tracking in \acs{IRS}-Assisted \acs{MIMO} Systems}
\author{Kenneth B. A. Benício, André L. F. de Almeida,~\IEEEmembership{Senior~Member,~IEEE}, Bruno Sokal, Fazal-E-Asim,~\IEEEmembership{Senior~Member,~IEEE}, Behrooz Makki,~\IEEEmembership{Senior~Member,~IEEE}, and Gábor Fodor,~\IEEEmembership{Senior~Member,~IEEE}

\thanks{This work was supported by the Ericsson Research, Sweden,
and Ericsson Innovation Center, Brazil, under UFC.51 Technical Cooperation Contract Ericsson/UFC. This study was financed in part by CAPES/Brazil - Finance Code 001, and CAPES/PRINT Proc. 88887.311965/2018-00. André L. F. de Almeida thanks CNPq for its financial support under grant 312491/2020-4. G. Fodor was partially
supported by the Digital Futures project PERCy.}}
    
\maketitle
 
\begin{abstract}
    This letter proposes a model for symbol detection in the uplink of \ac{IRS}-assisted networks in the presence of channel aging. During the first stage, we model the received pilot signal as a tensor, which serves as a basis for both estimating the channel and configuring the \acs{IRS}. In the second stage, the proposed tensor approach tracks the aging process to detect and estimate the transmitted data symbols. Our evaluations show that our proposed channel and symbol estimation schemes improve the performance of \acs{IRS}-assisted systems in terms of the achieved bit error rate and mean squared error of the received data, compared to state of the art schemes.
\end{abstract}

\begin{IEEEkeywords}
    channel aging, channel estimation, intelligent reflecting surfaces, tensor-based algorithm
\end{IEEEkeywords}

\section{Introduction}  
    Over the last few years, \ac{IRS} has been considered as one of the possible technologies to be deployed in \ac{B5G} wireless networks due to their potential to improve system capacity \cite{rajatheva2021scoring,di2020smart,zheng2022survey,wu2021intelligent}. An \ac{IRS} is a $2$D panel composed of many passive reflecting elements whose phase shifts are adjusted to maximize the \ac{SNR} at the intended receiver \cite{gong2020toward}. Hence, channel estimation must be performed at the end nodes of the network and the receiver should estimate the involved channels from received pilots reflected by the \ac{IRS} according to a training protocol. Several works have addressed this problem, e.g., \cite{de2021channel,de2022semi,pan2022overview}. As pointed out in \cite{pan2022overview}, channel estimation methods can be divided into  unstructured and structured methods, where the latter category exploits the parametric (geometric) modeling of the cascaded channel, which is the focus of the present work.
        
    The authors in \cite{de2021channel} use a tensor approach to perform supervised channel estimation. Then, \cite{de2022semi} proposes a tensor-based receiver formulated as a semi-blind problem that jointly estimates the involved channels and transmitted data. However, these two works assume (quasi-)static channels and do not consider the aging problem, which is likely to be present due to user mobility. Finally, it is worth noting that the impact of channel aging in \ac{MIMO} and \ac{IRS}-assisted systems has been studied in \cite{fodor2021performance,makki2013feedback,cao2022two,9875036}  while the geometrical structure of the channel has not been exploited.
    
    
    In this letter,  we propose a signal modeling that exploits the geometric channel structure of the \ac{IRS}-assisted \ac{MIMO} network to estimate the spatial signatures of the network. 
    The time-varying fading coefficients are modeled by means of an \ac{AR} model with each channel changing independently at different time scales. Then, we formulate a two-stage tensor-based framework for parameter estimation and data-aided tracking.
    In the first stage, referred to as \ac{PARKRON}, the estimation of channel steering vectors (static parameters) is carried out by means of a constrained tensor-based solution. Then, in the second stage, referred to as \ac{TBT}, we perform data-aided tracking of channel fading coefficients and symbol detection. Simulation results show that our proposed method accurately tracks the cascaded channel while outperforming competing methods in terms of \ac{NMSE}.

    \textit{Notation}: Scalars, vectors, matrices, and tensors are represented by $a, \boldsymbol{a}, \boldsymbol{A}$, and $\mathcal{A}$. Also, $\boldsymbol{A}^{*}$, $\boldsymbol{A}^{\text{T}}$, $\boldsymbol{A}^{\text{H}}$, and $\boldsymbol{A}^{\dagger}$ stand for the conjugate, transpose, Hermitian, and pseudo-inverse, of a matrix $\boldsymbol{A}$. The $j$th column of $\boldsymbol{A} \in \mathbb{C}^{I \times J}$ is denoted by $\boldsymbol{a}_{j} \in \mathbb{C}^{I \times 1}$. The operator vec$(\cdot)$ transforms a matrix into a vector by stacking its  columns, e.g., $\text{vec}(\boldsymbol{A}) = \boldsymbol{a} \in \mathbb{C}^{IJ \times 1}$, while the unvec$(\cdot)_{{I \times J}}$ operator undo the operation. The operator D$(\cdot)$ converts a vector into a diagonal matrix,  $\text{D}_j(\boldsymbol{B})$ forms a diagonal matrix $R \times R$ out of the $j$th row of $\boldsymbol{B} \in \mathbb{C}^{J \times R}$. Also, $\boldsymbol{I}_{N}$ denotes an identity matrix of size $N \times N$. The symbols $\otimes$ and $\diamond$ indicate the Kronecker and Khatri-Rao products.

\section{System Model}
\begin{figure}[!t]
        \centering
        \includegraphics[scale = 0.575]{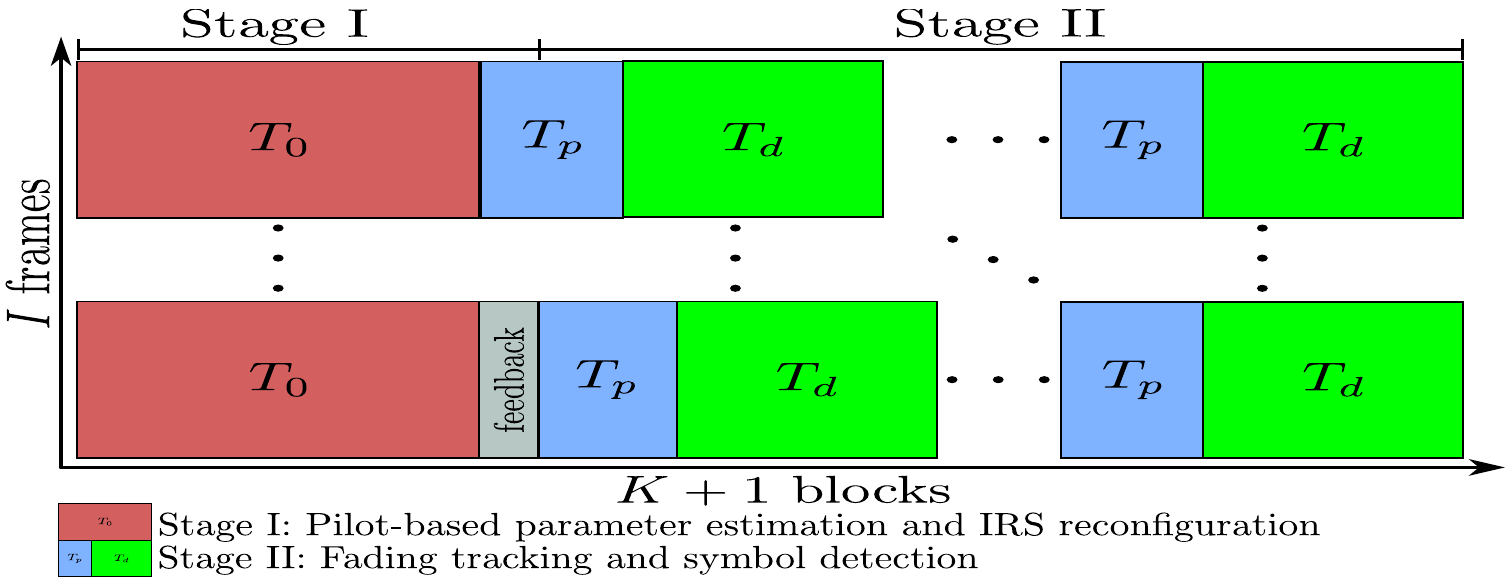}
        \caption{Time-domain transmission protocol.}
        \label{fig:01}
    \end{figure}

    We  consider an uplink \ac{IRS}-assisted \ac{MIMO} scenario with a \ac{BS} equipped with $M$ receiver antennas, which receives a signal from a \ac{UE} equipped with $Q$ transmit antennas \textit{via} a passive \ac{IRS} with $N$ reflecting elements. The transmission protocol is structured as $I$ frames each one containing $K+1$ blocks. The first block has length $T_{0}$ symbol periods, while the remaining $K$ blocks have length $(T_{\text{p}} + T_{\text{d}})$, as shown in Fig~\ref{fig:01}. The first block ($k=1$) of each frame is dedicated to pilot-aided parameter estimation, and the received signal is given by 
    \begin{align}
        \boldsymbol{y}_{i,1,t} &\hspace{-0.05cm}=\hspace{-0.05cm} \boldsymbol{G}_{i} \text{D}(\boldsymbol{s}_{t}) \boldsymbol{H}_{i,1} \boldsymbol{z}_{i,t} + \boldsymbol{v}_{i,1,t} \in \mathbb{C}^{M \times 1}, \label{signal_model_1}
    \end{align}
    where $\boldsymbol{z}_{i,t}$ is the pilot sequence and  $\boldsymbol{v}_{i,1,t}$ is the \ac{AWGN} vector with $t \in \{1, \cdots, T_{\text{0}}\}$. For the remaining $K$ blocks, the received signal can be written as
    \begin{align}
        \boldsymbol{Y}_{i,k} = \boldsymbol{G}_{i} \text{D}(\boldsymbol{s}_{\text{opt}}) \boldsymbol{H}_{i,k} \boldsymbol{X}_{i} + \boldsymbol{V}_{i,k} \in \mathbb{C}^{M \times (T_{\text{p}} + T_{\text{d}})} \label{signal_model_2}, 
    \end{align}
    where $\boldsymbol{Y}_{i,k} = [\boldsymbol{Y}^{\text{(p)}}_{i,k} | \boldsymbol{Y}^{(\text{d})}_{i,k}] \in \mathbb{C}^{M \times (T_{\text{p}} + T_{\text{d}})}$ and $\boldsymbol{X}_{i} = [\boldsymbol{X}^{\text{(p)}}_{i} | \boldsymbol{X}^{(\text{d})}_{i}] \in \mathbb{C}^{Q \times (T_{\text{p}} + T_{\text{d}})}$ are the received and the transmitted signals containing both pilots and data, spanning $T = T_{\text{p}} + T_{\text{d}}$ symbol periods each, where the pilot signals $\boldsymbol{Y}^{\text{(p)}}_{i,k} = [\boldsymbol{Y}^{\text{(p)}}_{i,k,1}, \cdots, \boldsymbol{Y}^{\text{(p)}}_{i,k,T_{\text{p}}}] \in \mathbb{C}^{M \times T_{\text{p}}}$ and $\boldsymbol{X}^{\text{(p)}}_{i} = [\boldsymbol{X}^{\text{(p)}}_{i,1}, \cdots, \boldsymbol{X}^{\text{(p)}}_{i,T_{\text{p}}}] \in \mathbb{C}^{Q \times T_{\text{p}}}$ have a duration of $T_{\text{p}}$ symbols and the data signals $\boldsymbol{Y}^{\text{(d)}}_{i,k} = [\boldsymbol{Y}^{\text{(d)}}_{i,k,1}, \cdots, \boldsymbol{Y}^{\text{(d)}}_{i,k,T_{\text{d}}}] \in \mathbb{C}^{M \times T_{\text{d}}}$ and $\boldsymbol{X}^{\text{(d)}}_{i} = [\boldsymbol{X}^{\text{(d)}}_{i,1}, \cdots, \boldsymbol{X}^{\text{(d)}}_{i,T_{\text{d}}}] \in \mathbb{C}^{M \times T_{\text{d}}}$ have a duration of $T_{\text{d}}$ symbols, respectively, in accordance to Fig. \ref{fig:01}. At the time instant $t$, $\text{D}(\boldsymbol{s}_{t})$ is the \ac{IRS} phase-shift matrix, $\text{D}(\boldsymbol{s}_{\text{opt}})$ the \ac{IRS} optimal phase-shift matrix obtained at the first block ($k = 1$), $\boldsymbol{X}^{\text{(p)}}_{i}$ is the pilot sequence, $\boldsymbol{X}^{\text{(d)}}_{i}$ is the data sequence.
    
    We assume that the \ac{IRS}-\ac{UE} link changes faster due to mobility while the \ac{BS}-\ac{IRS} link changes more slowly due to possible changes in interference. Specifically, the \ac{IRS}-\ac{UE} channel $\boldsymbol{H}_{i,k}$ changes between blocks within a given frame, while the \ac{BS}-\ac{IRS} channel $\boldsymbol{G}_{i}$ changes at a larger time scale, remaining constant during a frame of $K+1$ blocks while varying across the $I$ frames. Assuming a mmWave scenario,  we adopt a multipath channel model \cite{heath2016overview} for the involved channels. We can express these channel matrices as follows
   \begin{align}
        \boldsymbol{G}_{i} &= \boldsymbol{A}_{\text{r}_{\text{x}}} \boldsymbol{D} (\boldsymbol{\alpha}_{i}) \boldsymbol{B}^{\text{H}}_{\text{t}_{\text{x}}} \in \mathbb{C}^{M \times N}, \label{eq:G} \\
        \boldsymbol{H}_{i,k} &= \boldsymbol{B}_{\text{r}_{\text{x}}} \boldsymbol{D} (\boldsymbol{\beta}_{i,k}) \boldsymbol{A}_{\text{t}_{\text{x}}}^{\text{H}} \in \mathbb{C}^{N \times Q} \label{eq:H},
    \end{align}
    \noindent where $\boldsymbol{A}_{\text{r}_{\text{x}}}$ and $\boldsymbol{B}_{\text{r}_{\text{x}}}$ are the steering matrices defined as 
    \begin{align*}
        \boldsymbol{A}_{\text{r}_{\text{x}}} &= \left[\boldsymbol{a}_{\text{r}_{\text{x}}}(\mu^{(1)}_{\text{bs}}), \cdots, \boldsymbol{a}_{\text{r}_{\text{x}}}(\mu^{(L_{1})}_{\text{bs}}) \right] \in \mathbb{C}^{M \times L_{1}}, \\
        \boldsymbol{B}_{\text{r}_{\text{x}}} &= \left[\boldsymbol{b}^{(\text{irs})}_{\text{r}_{\text{x}}}(\mu^{(1)}_{\text{irs}_{{\text{A}}}},\psi^{(1)}_{\text{irs}_{\text{A}}}), \cdots, \boldsymbol{b}^{\text{(irs)}}_{\text{r}_{\text{x}}}(\mu^{(L_{2})}_{\text{irs}_{{\text{A}}}},\psi^{(L_{2})}_{\text{irs}_{\text{A}}}) \right] \in \mathbb{C}^{N \times L_{2}},
    \end{align*}
    \noindent with $\boldsymbol{B}_{\text{t}_{\text{x}}} $ and $\boldsymbol{A}_{\text{t}_{\text{x}}}$ being defined similarly. 
    The $l$th \ac{BS} steering vector $\boldsymbol{a}(\mu^{(l_{1})}_{\text{bs}})$ is associated with the spatial frequency  $\mu^{(l_{1})}_{\text{bs}} = \pi \text{cos}(\phi^{(l_{1})}_{\text{bs}})$, with $\phi^{(l_{1})}_{\text{bs}}$ being the \ac{AoA}, which can be further written as 
    \begin{align}
        \boldsymbol{a}_{\text{r}_{\text{x}}}(\mu^{(l_{1})}_{\text{bs}}) = \left[1, \cdots, e^{-j\pi (M - 1) \mu^{(l_{1})}_{\text{bs}}}\right]^{\text{T}} \in \mathbb{C}^{M \times 1}.
    \end{align}
Similarly, the $p$th one-dimensional steering vector for the \ac{UE} is $\boldsymbol{c}(\mu^{(l_{2})}_{\text{ue}})$ having spatial frequency, which is defined as $\mu^{(l_{2})}_{\text{ue}} = \pi \text{cos}(\phi^{(l_{2})}_{\text{ue}})$, 
    with $\phi^{(l_{2})}_{\text{ue}}$ being the \ac{AoD}, and can be written in terms of spatial frequency as 
    \begin{align}
        \boldsymbol{a}_{\text{t}_{\text{x}}}(\mu^{(l_{2})}_{\text{ue}}) = \left[1, \cdots, e^{-j\pi (Q - 1) \mu^{(l_{2})}_{\text{ue}}}\right]^{\text{T}} \in \mathbb{C}^{Q \times 1}.
    \end{align}
At the \ac{IRS}, $\boldsymbol{b}^{(\text{irs})}_{\text{r}_{\text{x}}}(\mu^{(l_{2})}_{\text{irs}_{{\text{A}}}},\psi^{(l_{2})}_{\text{irs}_{\text{A}}})$ is the 2D steering 
    vector with spatial frequencies $\mu^{(l_{2})}_{\text{irs}_{{\text{A}}}} = \pi \text{cos}(\phi^{(l_{2})}_{\text{irs}_{\text{A}}}) \text{sin} (\theta^{(l_{2})}_{\text{irs}_{\text{A}}})$
    and $\psi^{(l_{2})}_{\text{irs}_{\text{A}}} = \pi \text{cos}(\phi^{(l_{2})}_{\text{irs}_{\text{A}}})$, where $\phi^{(l_{2})}_{\text{irs}_{\text{A}}}$ and $\theta^{(l_{2})}_{\text{irs}_{\text{A}}}$ 
    are the azimuth \ac{AoA} and the elevation \ac{AoA}, respectively. This can be further written as the Kronecker product between two steering vectors as
    \begin{align}
        \boldsymbol{b}^{\text{(irs)}}_{\text{r}_{\text{x}}}(\mu^{(l_{2})}_{\text{irs}_{{\text{A}}}},\psi^{(l_{2})}_{\text{irs}_{\text{A}}}) = \boldsymbol{b}^{\text{(irs)}}_{\text{r}_{\text{x}}}(\mu^{(l_{2})}_{\text{irs}_{{\text{A}}}}) \otimes \boldsymbol{b}^{\text{(irs)}}_{\text{r}_{\text{x}}}(\psi^{(l_{2})}_{\text{irs}_{\text{A}}}) \in \mathbb{C}^{N \times 1}.
    \end{align}
The \ac{IRS} transmission steering vector, $\boldsymbol{b}^{\text{(irs)\text{H}}}_{\text{t}_{\text{x}}}(\mu^{(l_{1})}_{\text{irs}_{{\text{D}}}},\psi^{(l_{1})}_{\text{irs}_{\text{D}}})$, is defined similarly. 
The \ac{IRS} phase-shift vector is defined as $\boldsymbol{s}_{t} = \left[e^{j \theta_{1,t}}, \cdots, e^{j \theta_{N,t}}\right]^{\text{T}}  \in \mathbb{C}^{N \times 1}$, where $\theta_{n,t}$ is the phase-shift of the $n$th \ac{IRS} element at the $t$th time slot. Moreover, $\boldsymbol{\alpha}_{i} = [\alpha^{(1)}_{i}, \cdots, \alpha^{(L_{1})}_{i} ]^{\text{T}} \in \mathbb{C}^{L_{1} \times 1}$ and $\boldsymbol{\beta}_{i,k} = [\beta^{(1)}_{i,k}, \cdots, \beta^{(L_{2})}_{i,k}]^{\text{T}} \in \mathbb{C}^{L_{2} \times 1}$ collect the path loss and fading components of the \ac{BS}-\ac{IRS} and \ac{IRS}-\ac{UE} links, respectively. 

The aging effects are modeled by assuming that  $\boldsymbol{\alpha}_{i} \in \mathbb{C}^{L_{1} \times 1}$ and $\boldsymbol{\beta}_{i,k} \in \mathbb{C}^{L_{2} \times 1}$ vary according to first-order \ac{AR} processes defined as \cite{fodor2021performance}
    \begin{align}
        \boldsymbol{\alpha}_{i} &= \delta \boldsymbol{\alpha}_{i-1} + \boldsymbol{\zeta}_{i}, \,\,\, i=1, \ldots, I,\label{alpha} \\
        \boldsymbol{\beta}_{i,k}& = 
        \begin{cases}
            \lambda \boldsymbol{\beta}_{i - 1, K} + \boldsymbol{\xi}_{i,k}, &  k = 1, \\
            \lambda \boldsymbol{\beta}_{i, k - 1} + \boldsymbol{\xi}_{i,k}, &  k = 2, \cdots, K,
        \end{cases} \label{beta}
    \end{align}

\noindent where $\boldsymbol{\zeta}_{i} \sim \mathcal{CN}(\boldsymbol{0},(1 -\delta^{2}) \boldsymbol{I}_{L_{1}}) \in \mathbb{C}^{L_{1} \times 1}$ and $\boldsymbol{\xi}_{i,k} \sim \mathcal{CN}(\boldsymbol{0},(1 -\lambda^{2}) \boldsymbol{I}_{L_{2}}) \in \mathbb{C}^{L_{2} \times 1}$ are the \ac{AR} process noise term for the \ac{BS}-\ac{IRS} and \ac{IRS}-\ac{UE} links with $\delta$ and $\lambda$ being their correlation coefficients \cite{makki2013feedback}, respectively.
\section{Pilot-based Parameter Estimation}
    \indent In this section, we formulate the parameter estimation problem and describe the \ac{IRS} phase-shift configuration approach shown in Algorithm \ref{alg:01} (\ac{PARKRON}). Here, we estimate the channel parameters, i.e., the array steering matrices and the complex channel gains of the first block ($k=1$) from all frames.  
    \subsection{Tensor-Based Parameter Estimation}  
In this section, we formulate a tensor-based approach to estimate the channel parameters. Using $\text{vec}(\boldsymbol{A} \boldsymbol{B} \boldsymbol{C}) = (\boldsymbol{C}^{\text{T}} \otimes \boldsymbol{A}) \text{vec}(\boldsymbol{B})$ and $\text{vec}(\boldsymbol{A} \text{D}\left(\boldsymbol{b}\right) \boldsymbol{C}) = (\boldsymbol{C}^{\text{T}} \diamond \boldsymbol{A})\boldsymbol{b}$ in (\ref{signal_model_1}), yields
        \begin{align}
            \notag \boldsymbol{y}_{i,1,t} &= \text{vec }(\boldsymbol{I}_{M} \boldsymbol{G}_{i} \text{D}(\boldsymbol{s}_{t}) \boldsymbol{H}_{i,1} \boldsymbol{z}_{i,t}) + \boldsymbol{v}_{i,1,t} \in \mathbb{C}^{M \times 1}, \\
            \notag &= (\boldsymbol{s}^{\text{T}}_{t} \otimes \boldsymbol{z}^{\text{T}}_{i,t} \otimes \boldsymbol{I}_{M}) \text{vec}(\boldsymbol{H}^{\text{T}}_{i,1} \diamond \boldsymbol{G}_{i}) + \boldsymbol{v}_{i,1,t}.
        \end{align}
Collecting the signals during the $T_{\text{0}}$ symbol periods yields
        \begin{align}
            \notag \boldsymbol{y}_{i,1} &= \left[ \boldsymbol{y}_{i,1,1}^{\text{T}}, \cdots, \boldsymbol{y}_{i,1,T_{0}}^{\text{T}} \right]^{\text{T}}, \\
            \notag &= [(\boldsymbol{S} \diamond \boldsymbol{Z}_{i})^{\text{T}} \otimes \boldsymbol{I}_{M}] \text{vec}(\boldsymbol{H}^{\text{T}}_{i,1} \diamond \boldsymbol{G}_{i}) + \boldsymbol{v}_{i,1} \in \mathbb{C}^{M T_{\text{0}} \times 1}, \nonumber \\ 
           &= \boldsymbol{\Omega}_{i}\boldsymbol{u} + \boldsymbol{v}_{i,1} \in \mathbb{C}^{M T_{\text{0}} \times 1}, \label{eq:collect}
         \end{align}
        where $\boldsymbol{S} = \left[ \boldsymbol{s}_{1}, \cdots, \boldsymbol{s}_{T_{0}} \right] \in \mathbb{C}^{N \times T_{\text{0}}}$, $\boldsymbol{Z}_{i} = \left[ \boldsymbol{z}_{i,1}, \cdots, \boldsymbol{z}_{i,T_{0}} \right] \in \mathbb{C}^{Q \times T_{\text{0}}}$ are matrices collecting the \ac{IRS} phase-shifts and pilots, $\boldsymbol{\Omega}_{i} = (\boldsymbol{S} \diamond \boldsymbol{Z}_{i})^{\text{T}} \otimes \boldsymbol{I}_{M} \in \mathbb{C}^{M T_{0} \times M Q N}$,        
        $\boldsymbol{u} = \textrm{vec}(\boldsymbol{H}^{\text{T}}_{i,1} \diamond \boldsymbol{G}_{i}) \in \mathbb{C}^{M Q N \times 1}$, and $\boldsymbol{v}_{i,1} = \left[\boldsymbol{v}^{\text{T}}_{i,1,1}, \cdots, \boldsymbol{v}^{\text{T}}_{i,1,T_{0}} \right]^{\text{T}} \in \mathbb{C}^{M T_{\text{0}} \times 1}$ is the \ac{AWGN} noise term.        
        From (\ref{eq:collect}), we obtain the following \ac{LS} problem
        \begin{align}
            \boldsymbol{\hat{u}}_{i} = \underset{\boldsymbol{u}_{i}}{\text{arg min}} \left|\left| \boldsymbol{y}_{i,1} - \boldsymbol{\Omega}_{i} \boldsymbol{u}_{i} \right|\right|^2_{2}, \label{lss}
        \end{align}
        where the solution requires $T_{\text{0}} \geq Q N$ and is given by
        \begin{align}
            \boldsymbol{\hat{u}}_{i} &= \boldsymbol{\Omega}^{\dagger}_{i} \boldsymbol{y}_{i,1} \in \mathbb{C}^{MQN \times 1}. 
            \label{eq:03}
        \end{align}
Let us define $\boldsymbol{R}_{i}=\textrm{unvec}_{MQ \times N}(\boldsymbol{\hat{u}}_{i}) \approx \boldsymbol{H}^{\text{T}}_{i,1} \diamond \boldsymbol{G}_{i} \in \mathbb{C}^{MQ \times N}$. Using (\ref{eq:G}) and (\ref{eq:H}), while applying property $(\boldsymbol{AC}) \diamond (\boldsymbol{BD}) = (\boldsymbol{A} \otimes \boldsymbol{B})(\boldsymbol{C} \diamond \boldsymbol{D})$, we have
        \begin{align}
            \notag \boldsymbol{R}_{i} &\approx [\boldsymbol{A}_{\text{t}_{\text{x}}}^{*} \text{D} (\boldsymbol{\beta}_{i} ) \boldsymbol{B}^{\text{T}}_{\text{r}_{\text{x}}}] \diamond [\boldsymbol{A}_{\text{r}_{\text{x}}}\text{D} (\boldsymbol{\alpha}_{i}) \boldsymbol{B}^{\text{H}}_{\text{t}_{\text{x}}}], \\
            &\approx \hspace{-0.075cm} (\hspace{-0.030cm} \boldsymbol{A}_{\text{t}_{\text{x}}}^{*} \hspace{-0.075cm} \otimes \hspace{-0.075cm} \boldsymbol{A}_{\text{r}_{\text{x}}}\hspace{-0.030cm} ) \hspace{-0.05cm} [\text{D} (\hspace{-0.025cm}\boldsymbol{\beta}_{i}\hspace{-0.025cm}) 
            \hspace{-0.075cm} \otimes \hspace{-0.075cm} \text{D} (\hspace{-0.025cm}\boldsymbol{\alpha}_{i}\hspace{-0.025cm})] 
            (\boldsymbol{B}^{\text{T}}_{\text{r}_{\text{x}}} \hspace{-0.075cm} \diamond \hspace{-0.075cm} \boldsymbol{B}^{\text{H}}_{\text{t}_{\text{x}}}). \label{eq:04}
        \end{align}
        Defining $\boldsymbol{F} = \left[\boldsymbol{f}_{1}, \cdots, \boldsymbol{f}_{I} \right]^\text{T} \in \mathbb{C}^{I \times L_{1} L_{2}}$
        with $\boldsymbol{f}_{i} = \boldsymbol{\beta}_{i} \otimes \boldsymbol{\alpha}_{i} \in \mathbb{C}^{L_{1} L_{2} \times 1}$, (\ref{eq:04}) can be expressed as
        \begin{align}
            \boldsymbol{R}_{i} &\approx (\boldsymbol{A}_{\text{t}_{\text{x}}}^{*} \otimes \boldsymbol{A}_{\text{r}_{\text{x}}}) \text{D}_{i}(\boldsymbol{F}) \boldsymbol{P}^{\text{T}}_{B} \in \mathbb{C}^{M Q \times N}, \label{eq:05}
        \end{align} 
        \noindent where $\boldsymbol{P}_{B} = (\boldsymbol{B}^{\text{T}}_{\text{r}_{\text{x}}} \diamond \boldsymbol{B}^{\text{H}}_{\text{t}_{\text{x}}}) \in \mathbb{C}^{N \times L_{1} L_{2}}$ combines the \ac{IRS} transmit and receive steering matrices. 
        The collection of matrices $\{\boldsymbol{R}_{1}, \ldots, \boldsymbol{R}_{I}\}$, in (\ref{eq:05}) over all frames $i \in \{1,\ldots,I\}$ can be arranged as a fourth-way tensor $\mathcal{R} \in \mathbb{C}^{M \times Q \times N \times I }$, which can be expanded in terms of $n$-mode products as
        \begin{align}
            \mathcal{R} &\approx \mathcal{I}_{4,L_{1} L_{2}} \times_{1} \left( \boldsymbol{A}_{\text{r}_{\text{x}}} \boldsymbol{\Psi} \right) \times_{2} \left( \boldsymbol{A}^{*}_{\text{t}_{\text{x}}} \boldsymbol{\Phi} \right) \times_{3} \boldsymbol{P}_{B} \times_{4} \boldsymbol{F}, \label{tensor}
        \end{align}
        \noindent where $\boldsymbol{\Psi} = \boldsymbol{1}^{\text{T}}_{L_{1}} \otimes \boldsymbol{I}_{L_{2}} \in \mathbb{R}^{L_{2} \times L_{1} L_{2}}$ and $\boldsymbol{\Phi} = \boldsymbol{I}_{L_{1}} \otimes \boldsymbol{1}^{\text{T}}_{L_{2}} \in \mathbb{R}^{L_{1} \times L_{1} L_{2}}$ are constraints matrices. This tensor structure follows a constrained PARAFAC decomposition, which can also be interpreted as a constrained factor decomposition \cite{de2008constrained} or parallel profiles with linear dependencies decomposition \cite{stegeman2010uniqueness}. Assuming that the \ac{BS} and the \ac{IRS} have fixed and known locations, it is reasonable to consider that the angular information between the \ac{IRS} and the \ac{BS} is known, i.e., we assume the knowledge of the steering matrix $\boldsymbol{A}_{\text{r}_{\text{x}}}$. Consequently, the estimation of $\boldsymbol{A}_{\text{t}_{\text{x}}}$, $\boldsymbol{P}_{B}$ and $\boldsymbol{F}$ consists of solving the following problem
        \begin{equation}
            \hspace{-0.25cm} \left\{\hspace{-0.08cm}\hat{\boldsymbol{A}}_{\text{t}_{\text{x}}}, \hat{\boldsymbol{P}_{B}}, \hat{\boldsymbol{F}}\right\} \hspace{-0.05cm}=\hspace{-0.05cm} \underset{\boldsymbol{A}_{\text{t}_{\text{x}}}, \boldsymbol{P}_{B}, \boldsymbol{F}}{\text{arg min}} \left|\left|\hspace{-0.05cm} \begin{split} \mathcal{R} - \mathcal{I}_{4,L_{1} L_{2}} \times_{1} \left(\boldsymbol{A}_{\text{r}_{\text{x}}} \boldsymbol{\Psi} \right) \\ 
            \times_{2} \left(\boldsymbol{A}^{*}_{\text{t}_{\text{x}}} \boldsymbol{\Phi} \right) \times_{3} \boldsymbol{P}_{B} \times_{4} \boldsymbol{F}\end{split} \right|\right|^{2}_{\text{F}}, \label{alsa}
        \end{equation}
        which can be performed by means of the well-known \ac{ALS} algorithm (see \cite{kolda2009tensor,comon2009tensor} for details), which delivers estimates of the involved steering matrices up to scaling ambiguities provided that the necessary conditions $QNI\geq L_1L_2$, $QMI\geq L_1L_2$, and $QMN\geq L_1L_2$ are satisfied \cite{stegeman2010uniqueness}. These conditions are related to the uniqueness of the \ac{LS} estimates of $\boldsymbol{A}^{*}_{\text{t}_{\text{x}}}$, $\boldsymbol{P}_{B}$, and $\boldsymbol{F}$, respectively. The scaling ambiguities can be easily removed since the steering matrices have a Vandermonde structure.
 \subsection{Khatri-Rao and Kronecker Factorizations}
 To obtain individual estimates of the steering matrices $\boldsymbol{B}_{\text{r}_{\text{x}}}$ and $\boldsymbol{B}_{\text{t}_{\text{x}}}$, 
        as well as the fading coefficients $\boldsymbol{f}_{\beta_{i}}$ and $\boldsymbol{\alpha}_{i}$, \ac{KRF} and \ac{KF} procedures are applied by solving the following problems
        \begin{align}
            &\left\{ \hat{\boldsymbol{B}}_{\text{t}_{\text{x}}}, \hat{\boldsymbol{B}}_{\text{r}_{\text{x}}} \right\} = \underset{\hat{\boldsymbol{B}}_{\text{r}_{\text{x}}}, \hat{\boldsymbol{B}}_{\text{t}_{\text{x}}}}{\text{arg min}} \left| \hspace{-0.05cm} \left| \hat{\boldsymbol{P}}_{B} - \boldsymbol{B}^{\text{T}}_{\text{r}_{\text{x}}} \diamond \boldsymbol{B}^{\text{H}}_{\text{t}_{\text{x}}} \right| \hspace{-0.05cm} \right|^{2}_{\text{F}}, \label{eq:krf} \\
            &\left\{ \hspace{-0.05cm} \hat{\boldsymbol{\beta}_{i}}, \hat{\boldsymbol{\alpha}_{i}} \hspace{-0.05cm} \right\} \hspace{-0.05cm}=\hspace{-0.05cm} \underset{\boldsymbol{\beta}_{i}, \boldsymbol{\alpha}_{i}}{\text{arg min}} \left| \hspace{-0.05cm} \left| \hspace{-0.05cm} \hat{\boldsymbol{f}}_{i} \hspace{-0.05cm}-\hspace{-0.05cm} \boldsymbol{\beta}_{i} \hspace{-0.05cm}\otimes\hspace{-0.05cm} \boldsymbol{\alpha}_{i} \hspace{-0.05cm} \right| \hspace{-0.05cm} \right|^{2}_{\text{2}}, i \hspace{-0.05cm}\in\hspace{-0.05cm} \{1, \cdots, I\}, \label{eq:kf}
        \end{align}
         the solutions of which are obtained by the \ac{KRF} and \ac{KF} algorithms as in \cite{de2021channel} and \cite{de2022semi}, respectively. The estimated and true matrices are linked as $\tilde{\boldsymbol{B}}_{\text{t}_{\text{x}}} = \boldsymbol{B}_{\text{t}_{\text{x}}} \Delta_{\boldsymbol{B}_{\text{t}_{\text{x}}}}$, $\tilde{\boldsymbol{B}}_{\text{r}_{\text{x}}} = \boldsymbol{B}_{\text{r}_{\text{x}}} \Delta_{\boldsymbol{B}_{\text{r}_{\text{x}}}}$, with 
 $\Delta_{\boldsymbol{B}_{\text{t}_{\text{x}}}} \Delta_{\boldsymbol{B}_{\text{r}_{\text{x}}}} = \boldsymbol{I}_{L_{1} L_{2}}$. Note that since these matrices have a Vandermonde structure, these scaling ambiguities can be removed by simple column normalization.
    \subsubsection*{\ac{IRS} phase-shift configuration} Upon estimation of the static channel parameters, \ac{IRS} configuration is accomplished. Let $\hat{\boldsymbol{R}}_{1} = (\hat{\boldsymbol{A}}_{\text{t}_{\text{x}}}^{*} \otimes \boldsymbol{A}_{\text{r}_{\text{x}}}) \text{D}_{1}(\hat{\boldsymbol{F}}) (\hat{\boldsymbol{B}}^{\text{T}}_{\text{r}_{\text{x}}} \diamond \hat{\boldsymbol{B}}^{\text{H}}_{\text{t}_{\text{x}}})^{\text{T}}$ be the reconstructed version of the combined channel in the first frame. The vector $\boldsymbol{s}_{\text{\text{opt}}} = \left[e^{j \theta_{1,\text{opt}}}, \cdots, e^{j \theta_{N,\text{opt}}}\right]^{\text{T}}$ containing the configured IRS phase shifts can be found from the dominant right singular vector of  $\boldsymbol{R}_{1} = \boldsymbol{U} \boldsymbol{\Sigma} \boldsymbol{V}^{\text{H}}$, which gives $\boldsymbol{s}_{\text{\text{opt}}} = e^{-j \angle{\boldsymbol{v}^{*}_{1}}}$.
           \begin{algorithm}[!t]
            \caption{Stage 1 (\acs{PARKRON}) \label{alg:01}}
            \begin{algorithmic}[1]
                \State{Transmit pilot signals with (\ref{signal_model_1}).}
                \State{Estimate $\notag \hat{\boldsymbol{u}}_{i} = \boldsymbol{\Omega}^{\dagger} \boldsymbol{y}_{i} \in \mathbb{C}^{MQN \times 1}.$}
                \State{Build $\boldsymbol{R}_{i} = \text{unvec}_{MQ \times N}(\boldsymbol{\hat{u}}_{i})$.}
                \State{From $\{\boldsymbol{R}_{1}, \ldots, \boldsymbol{R}_{I}\}$, build the tensor $\mathcal{R}$ in (\ref{tensor})}.
                \State{Estimate $\hat{\boldsymbol{A}}_{\text{t}_{\text{x}}}$, $\hat{\boldsymbol{P}}_{B}$, and $\hat{\boldsymbol{F}}$ by solving (\ref{alsa}) using \acs{ALS}.}
                \State{Estimate $\hat{\boldsymbol{B}}^{\text{T}}_{\text{r}_{\text{x}}}$ and $\hat{\boldsymbol{B}}^{\text{H}}_{\text{t}_{\text{x}}}$ by solving (\ref{eq:krf}) using \acs{KRF}.}
                \State{Estimate $\hat{\boldsymbol{F}}_{\alpha}$ and $\hat{\boldsymbol{F}}_{\beta}$ by solving (\ref{eq:kf}) using \acs{KF}.}
                \State{Rebuild $\hat{\boldsymbol{R}}_{1}=(\hat{\boldsymbol{A}}_{\text{t}_{\text{x}}}^{*} \otimes \boldsymbol{A}_{\text{r}_{\text{x}}}) \text{D}_{1}(\hat{\boldsymbol{F}})\boldsymbol{\hat{P}}^{\text{T}}_{B}$ and configure the \ac{IRS} phase shifts from its dominant right singular vector.}
            \end{algorithmic}
        \end{algorithm}
        \begin{algorithm}[!t]
            \caption{Stage 2 (\acs{TBT}) \label{alg:02}}
            \begin{algorithmic}[1]
                \State{Build the pilot tensor $\mathcal{Y}^{(\text{p})}_{i}$ in (\ref{received_pilot_tensor})}
                \State{Obtain an initial estimate  $\hat{\boldsymbol{F}}_{i}$ according to (\ref{fading_tracked}).}
                \State{Build the data tensor $\mathcal{Y}^{(\text{d})}_{i}$ in (\ref{received_data_tensor})}.
                \State{Estimate $\hat{\boldsymbol{X}}^{(\text{d})}_{i}$ and refine $\hat{\boldsymbol{F}}_{i}$ by solving (\ref{bals}) using \acs{BALS}.}
            \end{algorithmic}
        \end{algorithm}

\section{Channel tracking and symbol detection}
Under aging effects, the initial channel parameter estimates obtained in the first stage can quickly become outdated. This means that a further procedure is needed to track changes due to aging, if we intend to detect the transmitted symbols. To this end, we formulate the second stage of the proposed receiver, which is dedicated to channel tracking and symbol detection based on the estimated steering matrices in the first stage. 

    \subsection{Initialization}
        From the pilot part of the received signal in (\ref{signal_model_2}), we have
        \begin{align}
            \boldsymbol{Y}^{(\text{p})}_{i,k} &= \boldsymbol{A}_{\text{r}_{\text{x}}} \text{D}(\boldsymbol{\alpha}_{i}) \boldsymbol{J} \text{D}(\boldsymbol{\beta}_{i,k}) \boldsymbol{C}^{\text{T}}_{i} + \boldsymbol{V}^{(\text{p})}_{i,k} \in \mathbb{C}^{M \times T_{\text{p}}},\nonumber
        \end{align}
        where $\boldsymbol{J} = \boldsymbol{B}^{\text{H}}_{\text{t}_{\text{x}}} \text{D}(s_{\text{\text{opt}}}) \boldsymbol{B}_{\text{r}_{\text{x}}}$, $\boldsymbol{C}^{\text{T}}_{i} = \boldsymbol{A}_{\text{t}_{\text{x}}}^{\text{H}} \boldsymbol{X}^{(\text{p})}_{i}$, and $\boldsymbol{V}^{(\text{p})}_{i,k}$ is the associated \ac{AWGN} component. Defining 
        $\boldsymbol{y}^{(\text{p})}_{i,k}=\text{vec}(\boldsymbol{Y}^{(\text{p})}_{i,k})$ and using the equivalence $\left(\boldsymbol{a}^{\text{T}} \diamond \boldsymbol{B}\right) = \boldsymbol{B} \text{D}(\boldsymbol{a})$ yields
        \begin{align}
            \notag \boldsymbol{y}^{(\text{p})}_{i,k} &= \left(\boldsymbol{C}_{i} \otimes \boldsymbol{A}_{\text{r}_{\text{x}}}\right) \text{vec}(\text{D}(\boldsymbol{\alpha}_{i}) \boldsymbol{J} \text{D}(\boldsymbol{\beta}_{i,k})) \in \mathbb{C}^{M T_{\text{p}} \times 1}, \\ 
            \notag \hspace{-0.05cm}&=\hspace{-0.05cm} \left(\boldsymbol{C}_{i} \hspace{-0.05cm}\otimes\hspace{-0.05cm} \boldsymbol{A}_{\text{r}_{\text{x}}}\right) \hspace{-0.05cm} \text{D}(\text{vec}(\boldsymbol{J})) \hspace{-0.05cm} \left(\boldsymbol{\beta}_{i,k} \hspace{-0.05cm}\otimes\hspace{-0.05cm} \boldsymbol{\alpha}_{i}\right)  + \boldsymbol{v}^{(\text{p})}_{i,k}.
        \end{align}
        Then, defining $\boldsymbol{Y}^{(\text{p})}_{i}=[\boldsymbol{y}^{(\text{p})}_{i,1}, \ldots, \boldsymbol{y}^{(\text{p})}_{i,K}] \in \mathbb{C}^{M T_{\text{p}} \times K}$ collecting all remaining $K$ blocks, we have
        \begin{align}
            \hspace{-0.15cm} \boldsymbol{Y}^{(\text{p})}_{i} \hspace{-0.05cm} &= \hspace{-0.05cm} \left(\boldsymbol{C}_{i} \hspace{-0.05cm}\otimes\hspace{-0.05cm} \boldsymbol{A}_{\text{r}_{\text{x}}}\right) \hspace{-0.05cm} \text{D}(\text{vec}(\boldsymbol{J})) \hspace{-0.05cm} \left(\boldsymbol{F}_{\beta_{i}} \hspace{-0.05cm}\diamond\hspace{-0.05cm} \boldsymbol{F}_{\alpha_{i}}\right)^{\text{T}} + \boldsymbol{V}^{(\text{p})}_{i},\nonumber
        \end{align}
        where $\boldsymbol{F}_{\alpha_{i}} = \boldsymbol{1}^{\text{T}}_{K} \otimes \boldsymbol{\alpha}_{i} \in \mathbb{C}^{L_{1} \times K}$, $\boldsymbol{F}_{\beta_{i}} = [\boldsymbol{\beta}_{i,2}, \cdots, \boldsymbol{\beta}_{i,K+1}] \in \mathbb{C}^{L_{2} \times K}$, and $\boldsymbol{V}^{(\text{p})}_{i}$ is the corresponding noise term. Note that $\text{D}(\text{vec}(\boldsymbol{J})) \in \mathbb{C}^{L_{1} L_{2} \times L_{1} L_{2}}$ can be viewed as the $3$-mode unfolding of the tensor 
        $\mathcal{J} \in \mathbb{C}^{L_{1} \times L_{2} \times L_{1} L_{2}}$ written as \cite{sokal2021tensor}
        \begin{align}
            \mathcal{J} &= \left(\mathcal{I}_{3,L_{2}} \otimes^{2,3}_{2,3} \mathcal{I}_{3,L_{1}}\right) \times_{3} \text{vec}(\boldsymbol{J}) \in \mathbb{C}^{L_{1} \times L_{2} \times L_{1} L_{2}},\nonumber
        \end{align}
        where $\mathcal{I}_{3,L_{2}}$ and $\mathcal{I}_{3,L_{1}}$ are identity tensors and $\otimes^{2,3}_{2,3}$ is the selective Kronecker product (SKP) \cite{sokal2021tensor}, from which we obtain
        \begin{align}
            \mathcal{Y}^{(\text{p})}_{i} &= \mathcal{J} \times_{1} \boldsymbol{A}_{\text{r}_{\text{x}}} \times_{2} \boldsymbol{C}_{i} \times_{3} \boldsymbol{F}^{\text{T}}_{i} +  \mathcal{V}^{(\text{p})}_{i} \in \mathbb{C}^{M \times T_{\text{p}} \times K}, \label{received_pilot_tensor}
        \end{align}
        where $\boldsymbol{F}_{i} = \boldsymbol{F}_{\beta_{i}} \hspace{-0.05cm}\diamond\hspace{-0.05cm} \boldsymbol{F}_{\alpha_{i}} \in \mathbb{C}^{L_{1} L_{2} \times K}$ is the combined fading matrix. The $3$-mode unfolding of the tensor is defined as
        \begin{align}
            [\mathcal{Y}^{(\text{p})}_{i}]_{(3)} &= \boldsymbol{F}^{\text{T}}_{i} \left[\mathcal{J}\right]_{(3)} (\boldsymbol{C}_{i} \otimes \boldsymbol{A}_{\text{r}_{\text{x}}})^{\text{T}} \in \mathbb{C}^{K \times M T_{\text{p}}},\nonumber
        \end{align}
        from which an \ac{LS} estimate of $\boldsymbol{F}_{i}$ can be obtained as
        \begin{align}
            \hspace{-0.15cm} \hat{\boldsymbol{F}}_{i} \hspace{-0.05cm}&=\hspace{-0.05cm} \Big[[\mathcal{Y}^{(\text{p})}_{i}]_{(3)} \Big([\hat{\mathcal{J}}]_{(3)} (\boldsymbol{C}_{i} \hspace{-0.05cm}\otimes\hspace{-0.05cm} \boldsymbol{A}_{\text{r}_{\text{x}}})^{\text{T}}\Big)^{\dagger}\Big]^{\text{T}} \hspace{-0.05cm}\in\hspace{-0.05cm} \mathbb{C}^{K \times L_{1} L_{2}}. \label{fading_tracked}
        \end{align}
        This initial LS step requires that $\left[\mathcal{J}\right]_{(3)} (\boldsymbol{C}_{i} \otimes \boldsymbol{A}_{\text{r}_{\text{x}}})^{\text{T}}\in \mathbb{C}^{L_{1} L_{2} \times M T_{\text{p}}}$ have full row rank, which implies $M T_{\text{p}} \geq L P$.
        \begin{figure}[!t]
            \centering
            \includegraphics[scale = 0.550]{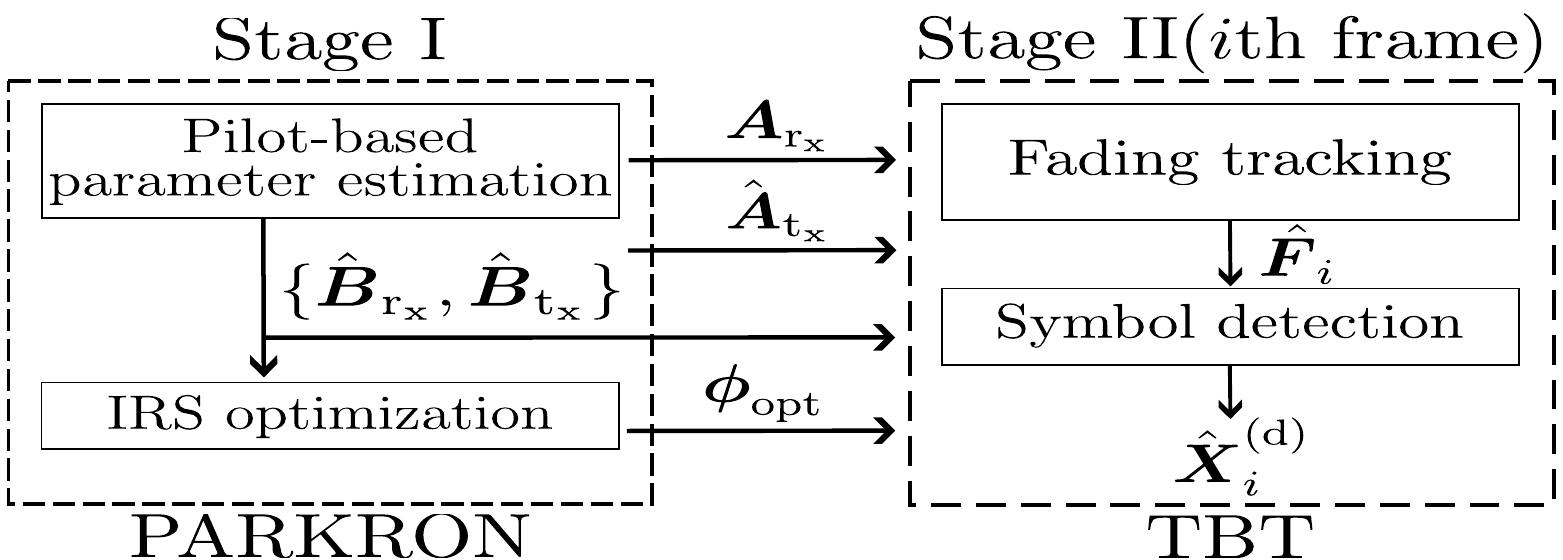}
            \caption{Block-diagram of the proposed receiver.}
            \label{fig:02}
        \end{figure}
    \subsection{Joint Tracking and Symbol Detection}
        \indent From the data part of the received signal in (\ref{signal_model_2}), we have
        \begin{align}
            \boldsymbol{Y}^{(\text{d})}_{i,k} &= \boldsymbol{A}_{\text{r}_{\text{x}}} \text{D}(\boldsymbol{\alpha}_{i}) \boldsymbol{J} \text{D}(\boldsymbol{\beta}_{i,k}) \boldsymbol{A}_{\text{t}_{\text{x}}}^{\text{H}} \boldsymbol{X}^{(\text{d})}_{i} + \boldsymbol{V}^{(\text{d})}_{i,k} \in \mathbb{C}^{M \times T_{\text{d}}},\nonumber
        \end{align}
        which can be reformulated in tensor form as
        \begin{align}
            \hspace{-0.25cm} \mathcal{Y}^{(\text{d})}_{i} &\hspace{-0.075cm}=\hspace{-0.075cm} \mathcal{J} \hspace{-0.075cm}\times_{1}\hspace{-0.075cm} \boldsymbol{A}_{\text{r}_{\text{x}}} \hspace{-0.075cm}\times_{2}\hspace{-0.075cm} (\boldsymbol{A}_{\text{t}_{\text{x}}}^{\text{H}} \boldsymbol{X}^{(\text{d})}_{i})^{\text{T}} \hspace{-0.075cm}\times_{3}\hspace{-0.05cm} \boldsymbol{F}^{\text{T}}_{i} \hspace{-0.05cm}+\hspace{-0.05cm} \mathcal{V}^{(\text{d})}_{i} \hspace{-0.05cm}\in\hspace{-0.05cm} \mathbb{C}^{M \times T_{\text{d}} \times K}\label{received_data_tensor}
        \end{align}
        The $2$-mode and $3$-mode unfoldings of $\mathcal{Y}^{(\text{d})}_{i}$ are given by
        \begin{align*}
            \left[\mathcal{Y}^{(\text{d})}_{i}\right]_{(2)} &= (\boldsymbol{A}_{\text{t}_{\text{x}}}^{\text{H}} \boldsymbol{X}^{(\text{d})}_{i})^{\text{T}} \left[\mathcal{J}\right]_{(2)} \left(\boldsymbol{F}^{\text{T}}_{i} \otimes \boldsymbol{A}_{\text{r}_{\text{x}}}\right)^{\text{T}} \in \mathbb{C}^{T_{\text{d}} \times M K}, \\
            \left[\mathcal{Y}^{(\text{d})}_{i}\right]_{(3)} &= \boldsymbol{F}^{\text{T}}_{i} \left[\mathcal{J}\right]_{(3)} \left[(\boldsymbol{A}_{\text{t}_{\text{x}}}^{\text{H}} \boldsymbol{X}^{(\text{d})}_{i})^{\text{T}} \otimes \boldsymbol{A}_{\text{r}_{\text{x}}}\right]^{\text{T}} \in \mathbb{C}^{K \times T_{\text{d}} M}.
        \end{align*}
        \indent Our joint tracking and symbol detection \ac{TBT} algorithm consists of estimating the data matrix $\boldsymbol{X}^{(\text{d})}_{i}$ while refining the estimate of $\boldsymbol{F}_{i}$ initialized with (\ref{fading_tracked}) by solving 
        \begin{equation}
            \hspace{-0.25cm} \left\{\hspace{-0.08cm} \hat{\boldsymbol{X}}^{(\text{d})}_{i}, \hat{\boldsymbol{F}}_{i}\right\} \hspace{-0.05cm}=\hspace{-0.05cm} \underset{\boldsymbol{X}^{(\text{d})}_{i}, \boldsymbol{F}_{i}}{\text{arg min}} \left|\left|\hspace{-0.05cm} \begin{split} \mathcal{Y}^{(\text{d})}_{i} - 
            \mathcal{J} \times_{1} \boldsymbol{A}_{\text{r}_{\text{x}}} \\ \times_{2} (\boldsymbol{A}_{\text{t}_{\text{x}}}^{\text{H}} \boldsymbol{X}^{(\text{d})}_{i})^{\text{T}} \times_{3} \boldsymbol{F}^{\text{T}}_{i} \end{split} \right|\right|^{2}_{\text{F}}, \label{bals}
        \end{equation}
        whose LS solutions are given by 
        \begin{align}
            \hat{\boldsymbol{X}}^{(\text{d})}_{i} \hspace{-0.05cm} &= \hspace{-0.05cm} \left[\left[\mathcal{Y}^{(\text{d})}_{i}\right]_{(2)} \hspace{-0.05cm}\left(\boldsymbol{A}_{\text{t}_{\text{x}}}^{*} \left[\mathcal{J}\right]_{(2)} (\hat{\boldsymbol{F}^{\text{T}}_{i}} \hspace{-0.05cm}\otimes\hspace{-0.05cm} \boldsymbol{A}_{\text{r}_{\text{x}}})^{\text{T}}\right)^{\dagger}\right]^{\text{T}}, \label{eq:xd} \\
            \hat{\boldsymbol{F}}_{i} \hspace{-0.05cm}&=\hspace{-0.05cm} \left[\left[\mathcal{Y}^{(\text{d})}_{i}\right]_{(3)} \left(\left[\mathcal{J}\right]_{(3)} \left(  (\boldsymbol{A}_{\text{t}_{\text{x}}}^{\text{H}} \boldsymbol{X}^{(\text{d})}_{i})^{\text{T}} \hspace{-0.05cm}\otimes\hspace{-0.05cm} \boldsymbol{A}_{\text{r}_{\text{x}}}\right)^{\text{T}}\right)^{\dagger}\right]^{\text{T}}. \label{eq:fihat}
        \end{align}
        The estimates of $\boldsymbol{X}^{(\text{d})}$ and $\boldsymbol{F}_{i}$ are obtained by alternating between the LS steps  (\ref{eq:xd}) and (\ref{eq:fihat}) using the \ac{BALS} algorithm \cite{de2021channel}, which requires $M K \geq Q$ and $M T_{\text{d}} \geq L_{1} L_{2}$. A summary of the channel tracking and symbol detection stage is shown in Algorithm \ref{alg:02}.  

         Table \ref{tab:complexity} contains a summary of the computational complexities of the proposed \ac{PARKRON} and \ac{TBT} algorithms, as well as that of the competing \ac{KRF} algorithm \cite{de2021channel} used as a benchmark. The terms $\text{\acs{ALS}}_{\text{iter}}$ and $\text{\acs{BALS}}_{\text{iter}}$ denote the number of iterations required for the convergence of the \ac{PARKRON} and \ac{TBT} algorithms, respectively. These results consider $\mathcal{O}(N_{1}N_{2})$ as the complexity associated with the rank-one approximation of a matrix $\mathbf{A} \in \mathbb{C}^{N_{1} \times N_{2}}$.
         Since the \ac{KRF} algorithm is a closed-form solution, it has lower computational complexity than the proposed receiver. However, note that KRF is limited to obtaining the unstructured estimates of $\boldsymbol{G}_{i}$ and $\boldsymbol{H}_{i,k}$, i.e., it does not provide estimates of the channel parameters since the associated channel structures are not exploited. A block diagram of the overall receiver processing, including the first and second stages. is shown in Fig. \ref{fig:02}.

\section{Simulation Results}
    \indent We evaluate the performance of the proposed tensor-based algorithm by comparing it with the reference parameter estimation method based on the \ac{KRF} \cite{de2021channel}. The  pilot signal matrix $\boldsymbol{X}^{(\text{p})}$ is designed as a Hadamard matrix and the data signal matrix $\boldsymbol{X}^{(\text{d})} $ follows a \ac{BPSK}, while a \ac{DFT} is adopted for the \ac{IRS} phase-shift matrix $\boldsymbol{S}$. The angular parameters $\phi^{(l_{1})}_{\text{bs}}$ and $\phi^{(l_{2})}_{\text{ue}}$ are randomly generated from a uniform distribution between $[-\pi, \pi]$ while the \ac{IRS} elevation and azimuth angles of arrival and departure are
    randomly generated from a uniform distribution between $[-\pi/2, \pi/2]$. The fading coefficients $\boldsymbol{\alpha}_{i}$ and $\boldsymbol{\beta}_{i,k}$ are 
    modeled as independent Gaussian random variables $\mathcal{CN}(0,1)$. The parameter estimation accuracy is evaluated in terms of the \ac{NMSE} given as $\text{\ac{NMSE}}(\boldsymbol{Q}) = \mathbb{E} \{\left|\left|\boldsymbol{Q}^{(e)} - \hat{\boldsymbol{Q}}^{(e)}\right|\right|^{2}_{\text{F}} / \left|\left|\boldsymbol{Q}^{(e)}\right|\right|^{2}_{\text{F}}\}$ with
    $\boldsymbol{Q} \in \{\boldsymbol{R}_{i}, \boldsymbol{W}_{i,k}\}$ being $\boldsymbol{R}_{i} = (\boldsymbol{H}^{\text{T}}_{i} \diamond \boldsymbol{G}_{i})$ and $\boldsymbol{W}_{i,k} = \boldsymbol{G}_{i} \text{diag}(\boldsymbol{s}_{\text{\text{opt}}})\boldsymbol{H}_{i,k}$ being used at the evaluation of the first stage and second stage at the $e$th run, $E = 10^4$ being the number of Monte Carlo runs. Symbol detection performance in the second stage is evaluated in terms of the bit error ratio (BER). Unless otherwise stated, the training \ac{SNR} is set to $30$ dB and the parameters are
    $\{M = 2, Q = 2, L_{1} = 2, L_{2} = 2, N = 32, T_{0} = T = 64, T_{\text{p}} = 16, T_{\text{d}} = 48, I = 2, K = 5, \lambda = 0.75, \text{ and } \delta = 0.75\}$. 
    
    
    \begin{table}[!t]
        \centering
        \caption{Computational complexity}.
        \label{tab:complexity}
        \resizebox{0.48\textwidth}{!}{%
        \begin{tabular}{|c|c|}
        \hline
        Algorithm   & Computational Complexity \\ \hline
        \acs{PARKRON}    &  $\mathcal{O}(L_{1} L_{2} (3\text{\acs{ALS}}_{\text{iter}}(L_{1}L_{2})^2 + N + I))$                         \\ \hline
        \acs{TBT}   &    $\mathcal{O}((L_{1} L_{2})^3 ( 1 + 2\text{\acs{BALS}}_{\text{iter}} ))$                 \\ \hline
        \acs{KRF} &      $\mathcal{O}(I K N L_{1} L_{2})$    \\ \hline
        \end{tabular}%
        }
    \end{table}
        
\begin{figure}[!t]
        \centering
        \begin{minipage}{.485\columnwidth}
            \centering
            \includegraphics[width = 0.95\linewidth, height =  4.9cm]{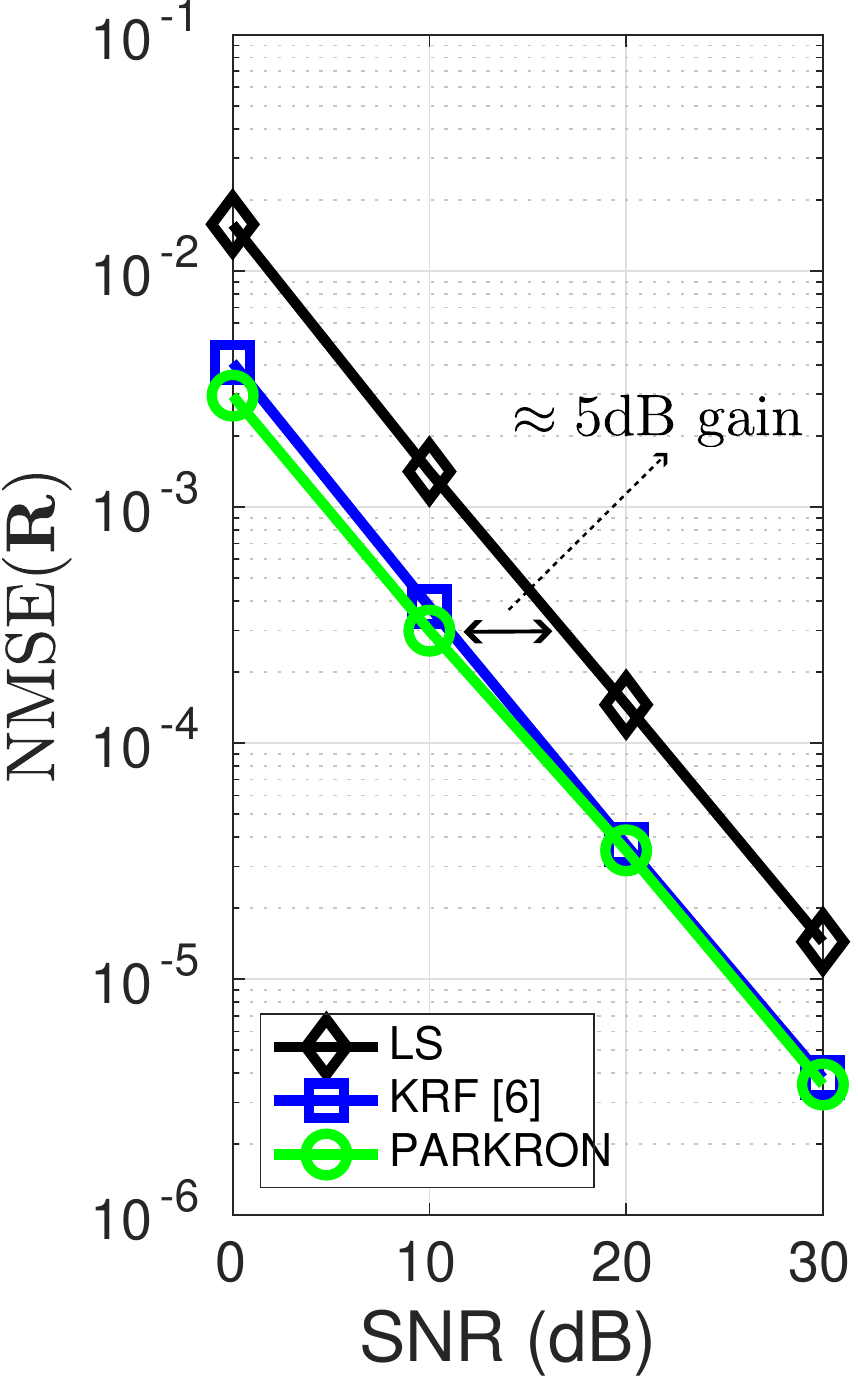}
            \caption{\ac{NMSE} at first stage.}
            \label{fig:03}
        \end{minipage}
        \begin{minipage}{.485\columnwidth}
            \centering
            \includegraphics[width = 0.95\linewidth, height = 4.9cm]{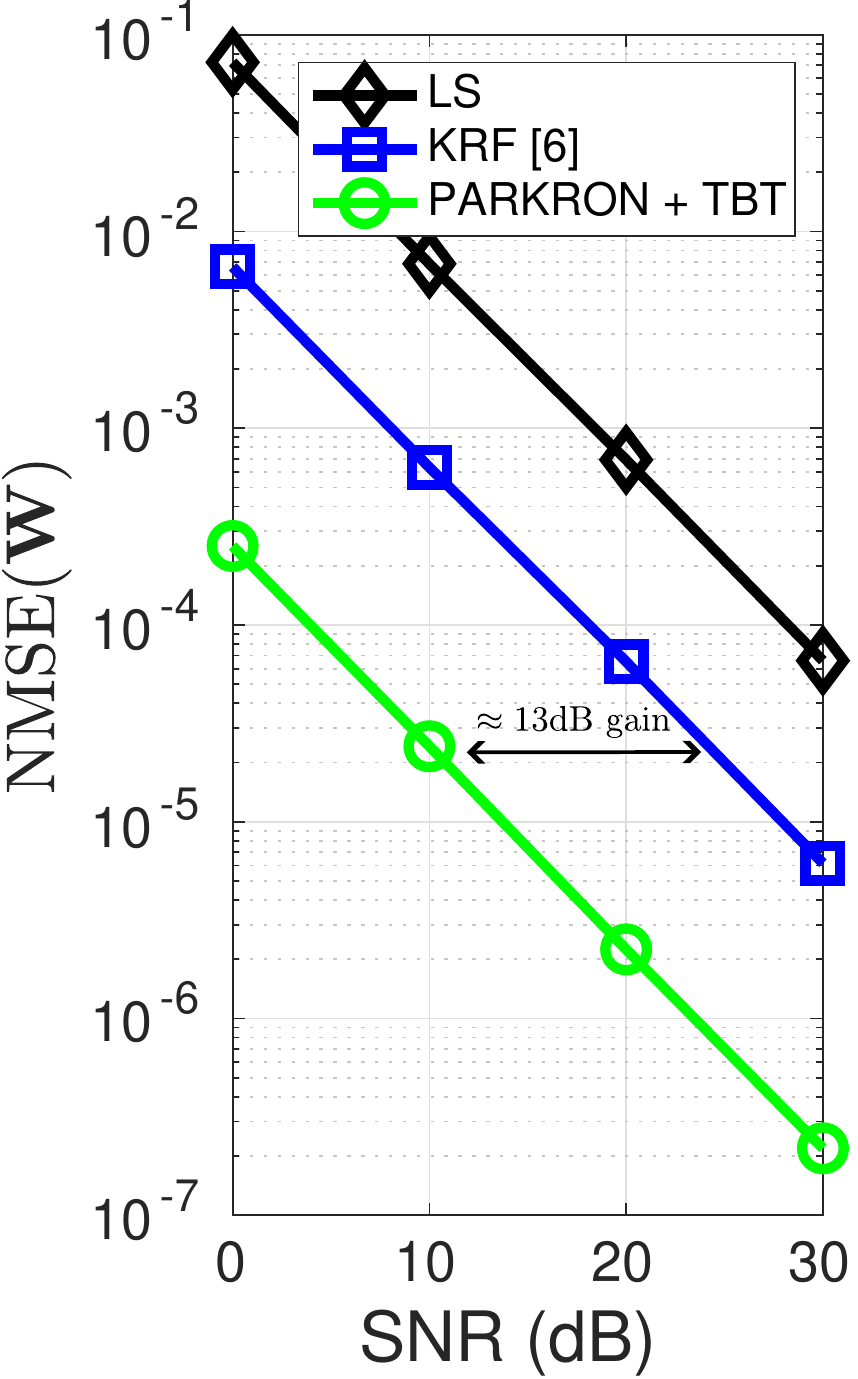}
            \caption{\ac{NMSE} at second stage.}
            \label{fig:04}
        \end{minipage}
    \end{figure}

In Fig. \ref{fig:03}, we compare the \ac{NMSE} performance of the proposed technique at the first stage. We take the \ac{NMSE} of $\boldsymbol{R}_{i}$ after the PARAFAC \ac{ALS} estimation and for comparison, we use the classical \ac{LS} and \ac{KRF} \cite{de2021channel} estimators. We notice that the \ac{PARKRON} and \ac{KRF} algorithms both outperform the \ac{LS} technique by approximately $5$ dB, almost independently of the \ac{SNR}. In Fig. \ref{fig:04}, we take the \ac{NMSE} of $\boldsymbol{W}_{i,k}$ after the \ac{TBT} algorithm and compare our method with the \ac{KRF} \cite{de2021channel}. We note that, for different \ac{SNR}s, the proposed \ac{PARKRON}-\ac{TBT} algorithm outperforms the \ac{KRF} \cite{de2021channel} with an approximate $13$ dB gain. This could be explained by how the proposed technique exploits the channel geometry while the \ac{KRF} \cite{de2021channel} does not take advantage of the approximately fixed geometry of the transmission. However, KRF \cite{de2021channel} is a closed-form solution having complexity $\mathcal{O}(I K N L_{1} L_{2})$. In comparison, our proposed framework is an iterative solution with the complexity of $\mathcal{O}(L_{1} L_{2}(3\text{\acs{ALS}}_{\text{iter}}(L_{1} L_{2})^{2} + N + I) + (L_{1}L_{2})^{3}(1 + 2\text{\acs{BALS}}_{\text{iter}}))$. Thus, our gains in \ac{NMSE} come at the cost of greater computational complexity.

\begin{figure}[!t]
        \centering
        \begin{minipage}{.485\columnwidth}
            \centering
            \includegraphics[width = 0.95\linewidth, height = 4.9cm]{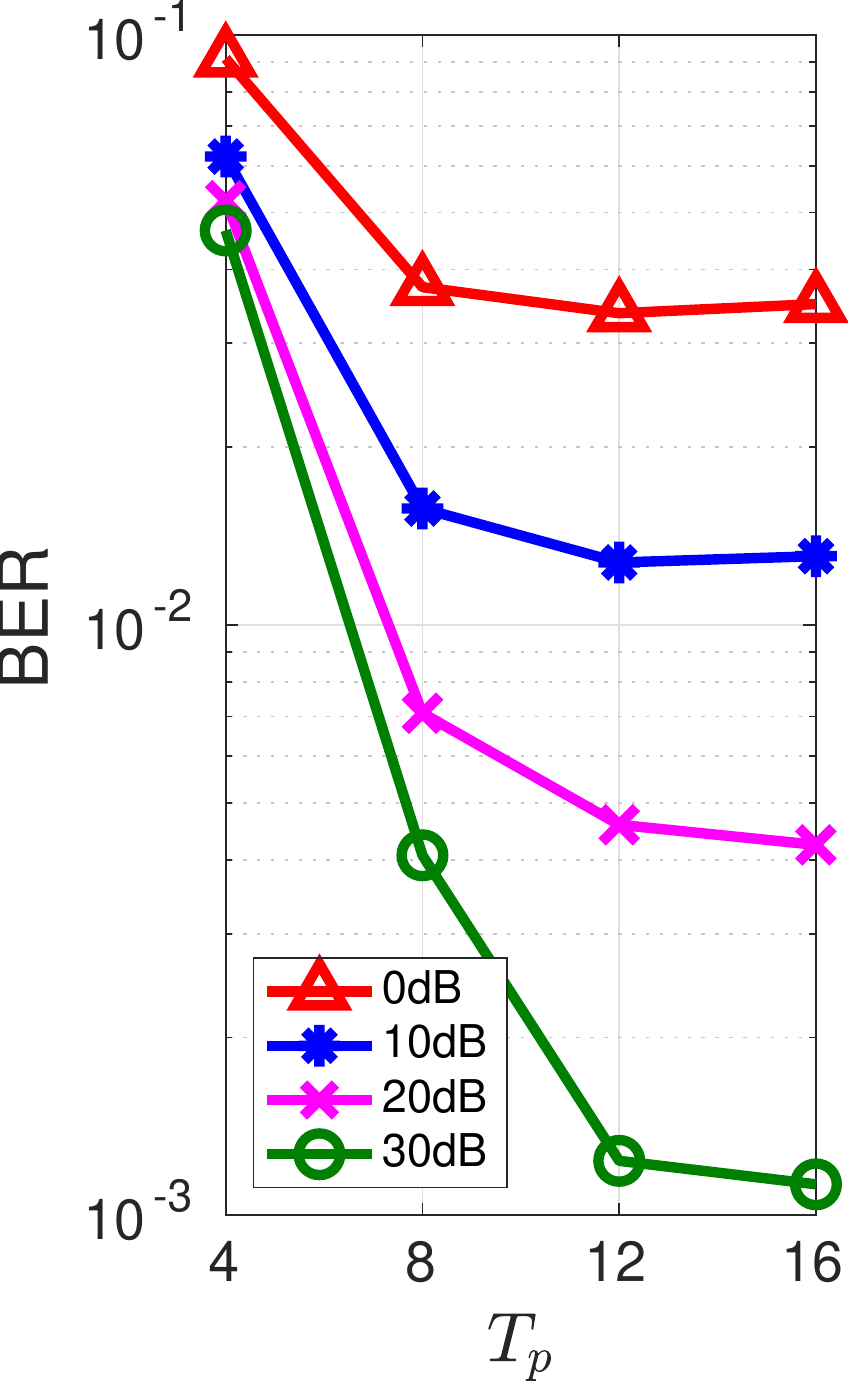}
            \caption{BER as a function \\ of number of pilot time-slots.}
            \label{fig:06}
        \end{minipage}
        \begin{minipage}{.485\columnwidth}
            \centering
            \includegraphics[width = 0.95\linewidth, height = 4.9cm]{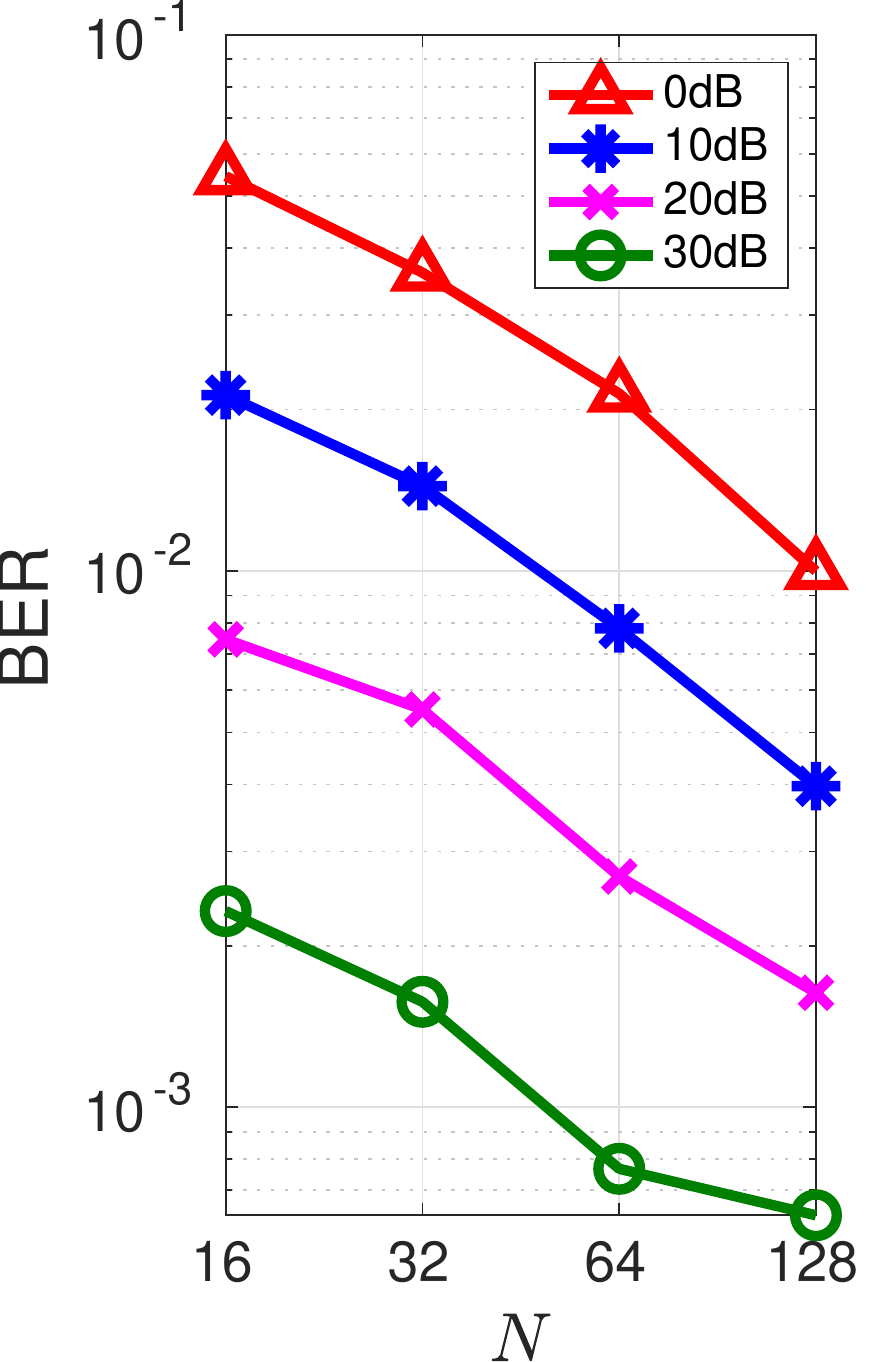}
            \caption{BER as a function of the number of reflecting elements.}
            \label{fig:07}
        \end{minipage}
    \end{figure}

In Fig. \ref{fig:06}, we study the effect of the  number of pilot-reserved time slots, $T_{\text{p}}$, on the BER of the proposed algorithm. We observe that the BER improves as a function of the number of pilots as well as the SNR, as expected. Since the size of each block is set to $T = 64$ the saturation point of each curve is close to $T_{\text{p}} = 16$ (or $25$\% of the available time-slots). In Fig. \ref{fig:07}, we evaluate the impact of the number $N$ of reflecting elements. Since the number of reflecting elements is directly linked to the size of the blocks at the first stage, $T_{0}$, as we increase $N$ we have better estimations of the combined channel parameters in (\ref{eq:03}) since we sense the channel longer. For the proposed scenario, the performance gains of increasing $N$, only to achieve better BER performance, seems too low to justify the increased algorithm complexity. 

\section{Conclusion}
    \indent We proposed a two-stage framework for channel parameter estimation, tracking and symbol detection for \ac{MIMO} \ac{IRS}-assisted communications under a double-channel aging model. The proposed scheme estimates the static parameters in the first stage to initialize a second stage dedicated to data-aided tracking of the aging process to estimate the transmitted data symbols. The proposed \ac{PARKRON}-\ac{TBT} framework efficiently exploits the higher-order tensor structure of the considered channel aging model, providing improved performance over competing channel estimation schemes that do not exploit the double aging structure for tracking purposes at the cost of higher computational complexity.


\end{document}